# Emergent Topological Superconductor by Charge Density Wave Transition

*Zishen Wang,[1,2] Jingyang You,[1] Chuan Chen,[3] Jinchao Mo,[1] Jingyu He,[1] Lishu Zhang,[1] Jun Zhou,[4,*] Kian Ping Loh,[2,5,*] Yuan Ping Feng[1,2,*]*

[1] Department of Physics, National University of Singapore, 117542 Singapore, Singapore

[2] Centre for Advanced 2D Materials, National University of Singapore, 117546 Singapore, Singapore

[3] Institute for Advanced Study, Tsinghua University, 100084 Beijing, China

[4] Institute of Materials Research & Engineering, A*STAR (Agency for Science, Technology and Research), 138634 Singapore, Singapore.

[5] Department of Chemistry, National University of Singapore, 117543 Singapore, Singapore

* Corresponding authors:

Jun Zhou (zhou_jun@imre.a-star.edu.sg)

Kian Ping Loh (chmlohkp@nus.edu.sg)

Yuan Ping Feng (phyfyp@nus.edu.sg)

**Abstract**

Many-body instabilities and topological physics are two attractive topics in condensed matter physics. It is intriguing to explore the interplay between these phenomena in a single quantum material. Here, using the prototypical charge density wave (CDW) material monolayer 1H-NbSe$_2$ as an example, we show how momentum-dependent electron-phonon coupling drives the CDW transition from $3 \times 3$ to $2 \times 2$ phase under electron doping. More interestingly, we find the coexistence of superconductivity and nontrivial topology in one of the two $2 \times 2$ CDW phases, the latter of which is identified by the nonzero Z$_2$ invariant with ideal Dirac cone edge states near the Fermi level. A similar CDW transition-induced topological superconductor has also been confirmed in monolayer 1H-TaSe$_2$. Our findings not only reveal a unique and general method to introduce nontrivial topology by CDW transition, but also provide an ideal platform to modulate different quantum orders by electron doping, thus stimulating experimental interest.

**Introduction**

Two-dimensional (2D) materials are among the most exciting research topics for their novel properties and promising applications in advanced electronic devices[1,2]. Specifically, transition metal dichalcogenides (TMDs) are an important member of the 2D material family, which offer an ideal platform to study quantum materials with topological physics and many-body instabilities including magnetism, superconductivity, and charge density wave (CDW)[3–6]. CDW is of particular interest as it can coexist or compete with other quantum phenomena. For example, CDW order and superconductivity usually compete with each other because the CDW gap opening decreases the density of states (DOS) at the Fermi level, which suppresses the formation of Cooper pairs below the superconducting transition temperature $(T_C)$[7,8]. Therefore, it is intriguing to modify electronic states at the Fermi level by external stimulus and study the interaction between CDW and superconductivity.

2H-NbSe$_2$ is a typical TMD that draws a great deal of attention due to the many-body instabilities[9–16]. The bulk 2H-NbSe$_2$ is a hole metal at room temperature, and it undergoes a CDW transition with in-plane $3 \times 3$ periodic distortion at $T_{CDW} \sim 33$ K as well as a superconducting transition at $T_C \sim 7$ K[9]. Both CDW and superconductivity have retained to the monolayer thickness[14–16]. With the decrease of the layer thickness, the suppression of the $T_C$ has been consistently demonstrated by many experiments[15–17]. However, the $T_{CDW}$ in the monolayer limit is under debate. The monolayer 1H-NbSe$_2$ obtained by molecular-beam epitaxy (MBE) has a low $T_{CDW} \sim 25$ K[15], while the samples obtained by the mechanical exfoliation exhibit a high $T_{CDW}$ ~ 145 K[14,18]. The discrepancy is believed to be due to the charge transfer from the substrates to the monolayer NbSe$_2$ samples[18–20].

In order to gain deeper insight into the relation between charge doping and the many-body instabilities of NbSe$_2$, a variety of doping methods have been used experimentally, including the electric double-layer transistor[17], ionic liquid gate[21], polymer electrolyte gating[22], Cu intercalation[23], and sandwiched NbSe$_2$ with an electron donor[24]. Interestingly, in contrast to the other CDW materials[25,26], both CDW and superconductivity in NbSe$_2$ are suppressed simultaneously under electron doping. Remarkably, NbSe$_2$ becomes a $2 \times 2$ CDW under high electron doping[24]. However, the basic understanding of this CDW transition and the physical properties of the $2 \times 2$ CDW structure are still absent.

In this work, based on first-principles calculations and mean-field theory, we perform a systematic investigation of the properties of monolayer NbSe$_2$ under electron doping. Momentum-dependent electron-phonon coupling (**q**-EPC) is found to account for both the $3 \times 3$ CDW formation and the phase transition from $3 \times 3$ to $2 \times 2$ CDW. Then we determine two possible $2 \times 2$ CDW structures for monolayer NbSe$_2$, which may exist when the doping concentration reaches 0.2 electrons per formula unit (e$^-$/f.u.). Besides, compared to the CDW gaps predicated by the mean-field model, the coupled electrons, which contribute mainly to the phonon softening, are found to be responsible for the CDW gap opening. In addition, the superconductivity of monolayer NbSe$_2$ is suppressed in the $2 \times 2$ CDW phases due to the reduced total electron-phonon coupling (EPC) strength under electron doping. More interestingly, we find one of the two $2 \times 2$ CDW phases topological nontrivial with a nonzero $Z_2$ invariant, corresponding to the Dirac cone edge states near the Fermi level. Thus, monolayer NbSe$_2$ can be an EPC-driven topological superconductor. Similar behavior is also observed in other TMDs, confirming its generality.

**A CDW Transition under Electron Doping**

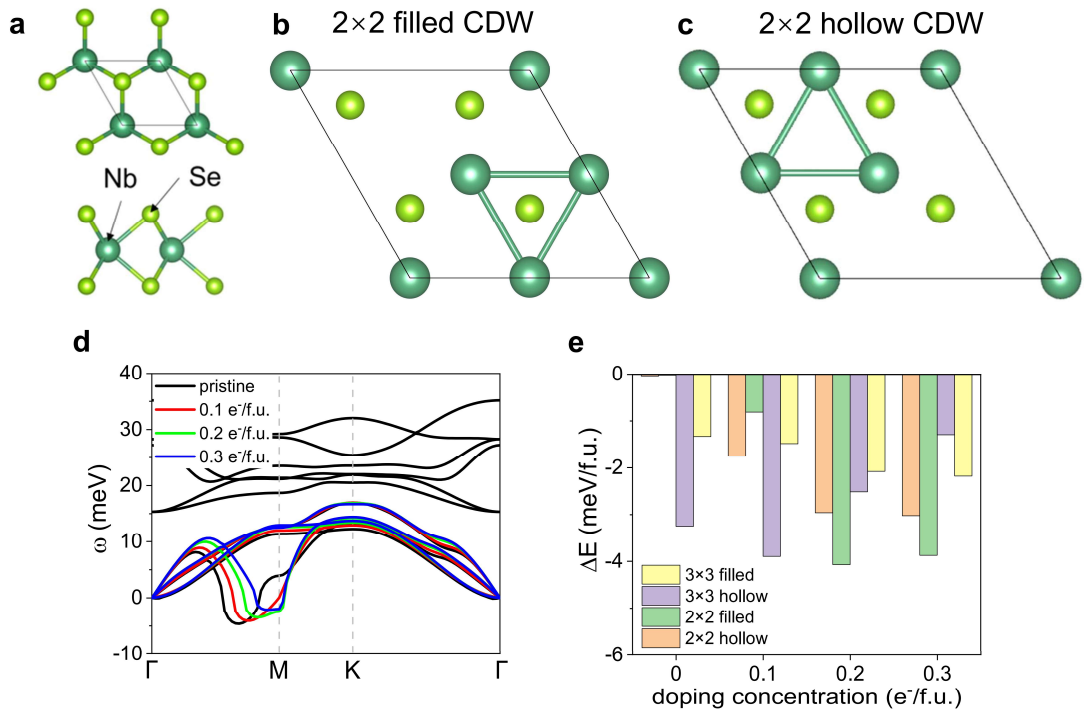

**Figure 1.** (a) Schematic diagram of normal NbSe$_2$ monolayer in the top and side views. (b) $2 \times 2$ filled and (c) $2 \times 2$ hollow NbSe$_2$ CDW structures. (d) The phonon spectra of NbSe$_2$ under different electron doping concentrations. Since the optical phonon modes hardly change, we only show acoustic phonon modes for systems under electron

doping (0.1-0.3 e⁻/f.u.). (e) The formation energies for the $3 \times 3$ filled, $3 \times 3$ hollow, $2 \times 2$ filled, and $2 \times 2$ hollow CDW structures as a function of electron doping concentration. The high-symmetry points are indicated in Figure S1.

The structure of normal monolayer NbSe₂ is shown in Figure 1a, which has a $P\bar{6}m2$ space group. Nb atoms are sandwiched between two Se layers with the trigonal prismatic cells. The optimized lattice constant of NbSe₂ is 3.474 Å, which is very close to the experimental value[15,19]. The phonon dispersion of normal NbSe₂ displays imaginary frequencies around $2/3\Gamma M$ (Figure 1d), corresponding to a $3 \times 3$ CDW. And it has been argued that there are two possible NbSe₂ $3 \times 3$ CDW structures (Figures. S2e and S2f), which are known as $3 \times 3$ hollow and $3 \times 3$ filled CDW structures with Nb clusters centered on the hollow sites and Se atoms of the honeycomb lattice, respectively[8,27,28]. With the increase of electron doping concentration, the position of imaginary phonons gradually evolves from $2/3\Gamma M$ to the $M$ point (Figure 1d), which implies a transition from $3 \times 3$ to $2 \times 2$ CDW. Similarly, we found two possible $2 \times 2$ CDW structures when doping concentration is above 0.2 e⁻/f.u., and we denote them as $2 \times 2$ filled (Figure 1b) and $2 \times 2$ hollow (Figure 1c) CDW structures according to the location of the centers of the Nb clusters.

To further verify CDW transition under electron doping in NbSe₂, the energetic hierarchy of these CDW structures is calculated at various doping concentrations. The CDW formation energies $\Delta E$ are calculated by $E_{CDW}/x - E_0$, where $E_{CDW}$ and $E_0$ are the total energies of the CDW and the $1 \times 1$ normal phases, respectively, $x$ is the number of formula units in the CDW phase. For the doping concentrations studied in this work, Figure 1f clearly shows a transition from $3 \times 3$ hollow CDW to $2 \times 2$ filled CDW at a critical doping concentration of around 0.2 e⁻/f.u., in agreement with phonon calculations (Figure 1d).

Next, we discuss the underlying physics of this CDW transition. It is known that the softened phonon frequency $\omega_\mathbf{q}$ can be obtained by[29]:

$$\omega_\mathbf{q}^2 = \Omega_\mathbf{q}^2 - 2\Omega_\mathbf{q}\Pi_\mathbf{q}, \tag{1}$$

where $\Omega_\mathbf{q}$ is the bare phonon frequency, and $\Pi_\mathbf{q}$ is the real part of the phonon self-energy, the latter of which carries the information from both EPC and Fermi surface nesting (FSN). In the static limit, $\Pi_\mathbf{q}$ has the form:

$$\Pi_\mathbf{q} = \sum_\mathbf{k} |g_{\mathbf{k},\mathbf{k}+\mathbf{q}}|^2 \frac{f(\varepsilon_\mathbf{k}) - f(\varepsilon_{\mathbf{k}+\mathbf{q}})}{\varepsilon_{\mathbf{k}+\mathbf{q}} - \varepsilon_\mathbf{k}}. \tag{2}$$

Here, $f(\varepsilon_\mathbf{k})$ is the Fermi-Dirac function with energy $\varepsilon$ at **k** point, and $g_{\mathbf{k},\mathbf{k}+\mathbf{q}}$ is the EPC matrix element which can be calculated via[30]:

$$g_{\mathbf{k},\mathbf{k}+\mathbf{q}} = \left(\frac{\hbar}{2M\omega_\mathbf{q}}\right)^{1/2} <\varphi_{\mathbf{k}+\mathbf{q}}|\partial_\mathbf{q} V|\varphi_\mathbf{k}>, \tag{3}$$

where $\varphi_\mathbf{k}$ is the electronic wavefunction with wavevector **k**, $V$ is the Kohn-Sham potential. The imaginary phonon frequency, suggesting CDW formation, appears when $\Pi_\mathbf{q}$ is larger than $(1/2)\Omega_\mathbf{q}$. The peak location of $\Pi_\mathbf{q}$ corresponds to the dip position of $\omega_\mathbf{q}$, which can be converted to the CDW vector in the reciprocal space, and the CDW lattice in the real space[31]. Due to the low energy physics studied in this work, the coupling is considered between the softened phonon branch and the single electronic band which crosses the Fermi level (Figures. 2a and 2c). Taking $|g_{\mathbf{k},\mathbf{k+q}}|^2 \sim |g_{\mathbf{q}}|^2$, which is known as **q**-EPC and can be extracted from $\Pi_{\mathbf{q}}$ to represent the EPC contribution[12], the remaining part is:

$$\chi_{\mathbf{q}} = \sum_{\mathbf{k}} \frac{f(\varepsilon_{\mathbf{k}}) - f(\varepsilon_{\mathbf{k+q}})}{\varepsilon_{\mathbf{k+q}} - \varepsilon_{\mathbf{k}}}, \qquad (4)$$

where $\chi_{\mathbf{q}}$ is the real part of the bare electronic susceptibility, which captures the Fermi surface topology as the FSN effect[32]. These quantities ($\Pi_{\mathbf{q}}$, $|g_{\mathbf{q}}|^2$ and $\chi_{\mathbf{q}}$) are further quantitatively analyzed to reveal the underlying mechanism for the NbSe$_2$ CDW formation and the phase transition from $3 \times 3$ to $2 \times 2$ CDW.

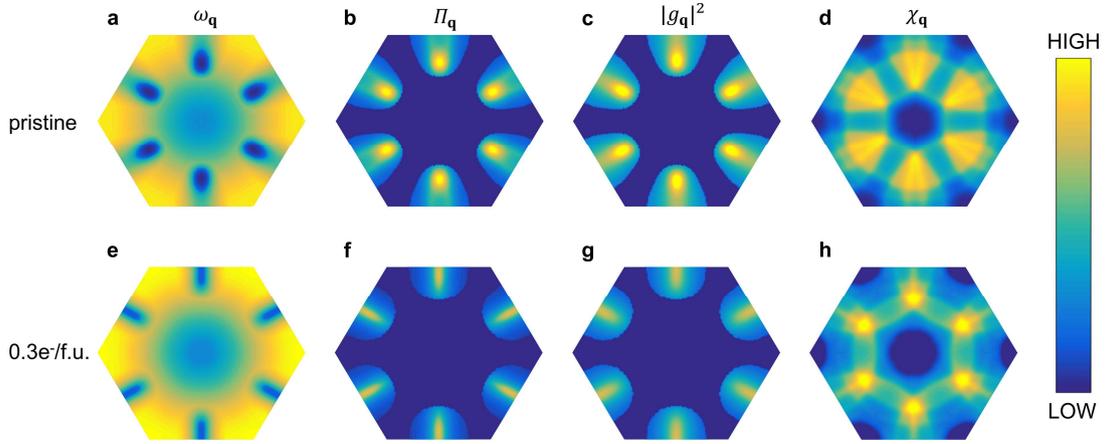

**Figure 2.** (a) The softened phonon frequency $\omega_{\mathbf{q}}$, (b) the real part of the phonon self-energy $\Pi_{\mathbf{q}}$, (c) **q**-EPC $|g_{\mathbf{q}}|^2$ and (d) the real part of the bare electronic susceptibility $\chi_{\mathbf{q}}$ in the first BZ of pristine NbSe$_2$. (e-h) Same as panels (a-d) but for NbSe$_2$ under 0.3 e$^-$/f.u. doping concentration.

Figures. 2a and 2e display the distribution of the softened phonon frequency $\omega_{\mathbf{q}}$ in the first Brillouin zone (BZ) of NbSe$_2$ without doping and under 0.3 e$^-$/f.u. doping concentration, respectively. The distribution of imaginary phonon frequencies (see the dark blue areas in Figures. 2a and 2e) clearly changes from the egg shape centered at $2/3\Gamma M$ to the valley shape centered at $M$ point. Meanwhile, a similar transition can be found for the peak of the real part of the phonon self-energy $\Pi_\mathbf{q}$ (see the yellow areas in Figures. 2b and 2f). Furthermore, we decompose the contribution of the real part of the phonon self-energy $\Pi_\mathbf{q}$ into pure effect of EPC and FSN by calculating $|g_\mathbf{q}|^2$ and $\chi_\mathbf{q}$ separately. For pristine NbSe$_2$, it is interesting to note that $|g_\mathbf{q}|^2$ and $\Pi_\mathbf{q}$ have very similar patterns in the first BZ, which both have maxima at $2/3\Gamma M$ (Figures. 2b and 2c). However, the $\chi_\mathbf{q}$ displays the highland in a fan-shaped region around $2/5\Gamma M$ (see the yellow region in Figure 2d), which deviates from the dip of $\omega_\mathbf{q}$ in Figure 2a and the peak of $\Pi_\mathbf{q}$ in Figure 2b. When the system is under 0.3 e$^-$/f.u. doping concentration, the distribution of $|g_\mathbf{q}|^2$ is again very close to $\Pi_\mathbf{q}$ with a ridge-like feature centered at $M$ point (Figures. 2f and 2g), corresponding to the valley in $\omega_\mathbf{q}$ (Figure 2e). The calculated $\chi_\mathbf{q}$ under 0.3 e$^-$/f.u. is shown in Figure 2h, and its peak at $1/2\Gamma M$ corresponds to a $4 \times 4$ CDW, in stark contrast with the experimental observation[24]. Thus, we can conclude that $|g_\mathbf{q}|^2$ provides the main contribution to $\Pi_\mathbf{q}$ and determines the phonon softening, confirming **q**-EPC as the driving force for the NbSe$_2$ CDW formation as well as the CDW transition from $3 \times 3$ to $2 \times 2$ CDW under electron doping.

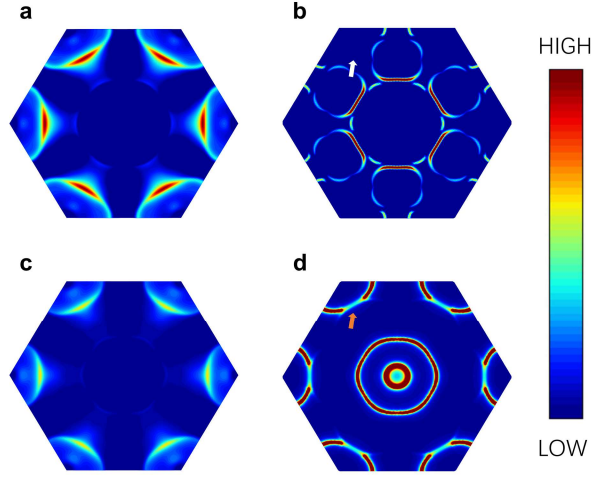

**Figure 3.** (a) The calculated $\Pi_\mathbf{k}$ in the first BZ of pristine NbSe$_2$. (b) The simulated Fermi surface of pristine NbSe$_2$ in ground-state $3 \times 3$ CDW phase. (c) The calculated $\Pi_\mathbf{k}$ in the first BZ of NbSe$_2$ under 0.3 e$^-$/f.u. doping concentration. (d) The simulated Fermi surface of NbSe$_2$ in ground-state $2 \times 2$ CDW phase under 0.3 e$^-$/f.u. doping concentration. The white arrow in (b) denotes the full CDW gap. The orange arrow in (d) denotes the partial CDW gap. The Fermi surfaces of normal phases of NbSe$_2$ are shown in Supplementary Figs. 2b and 2d.

To elucidate the **k**-resolved contributions to $\Pi_\mathbf{q}$, it is instructive to change the sum over **k** to the sum over **q**, which leads to:

$$\Pi_\mathbf{k} = \sum_\mathbf{q} |g_{\mathbf{k},\mathbf{k}+\mathbf{q}}|^2 \frac{f(\varepsilon_\mathbf{k}) - f(\varepsilon_{\mathbf{k}+\mathbf{q}})}{\varepsilon_{\mathbf{k}+\mathbf{q}} - \varepsilon_\mathbf{k}}. \quad (5)$$

Here, $\Pi_\mathbf{k}$ diagnostically indicates the distribution of the coupled electrons in the reciprocal space. For pristine NbSe$_2$, the coupled electrons are concentrated on the $K$ pocket, and they display the maximum intensity at the $\Gamma K$ path (see the red arc in Figure 3a). Interestingly, the high-intensity sector of $\Pi_\mathbf{k}$ is consistent with the full CDW gap position (see the white arrow in Figure 3b), as revealed by the calculated spectral function (see Supplementary information I for details). It is expected as these electrons are strongly disturbed by phonons. When NbSe$_2$ is under electron doping, the intraband screening processes are suppressed, which decreases the coupling between electrons and phonons. Thus, the peak of $\Pi_{\mathbf{k}}$ decreases and its position moves towards the $K$ point due to the shrink of the K pocket under the electron doping (see the yellow arc in Figure 3c). And the band-gapped sector in the $K$ pocket also decreases with a full to partial gap transition (see the orange arrow in Figure 3d).

**A 2D Topological Superconductor**

After studying the CDW properties of NbSe$_2$, we now explore its superconducting properties. According to the Allen-Dynes modified McMillan theory[30], the Eliashberg spectral function is obtained by:

$$\alpha^2 F(\omega) = \frac{1}{2N_{\mathbf{q}}} \sum_{\mathbf{q}\nu} \lambda_{\mathbf{q}\nu} \omega_{\mathbf{q}\nu} \delta(\omega - \omega_{\mathbf{q}\nu}), \tag{6}$$

where $\lambda_{\mathbf{q}\nu}$ is the electron-phonon coupling strength of phonon mode $\nu$ with momentum $\mathbf{q}$, $\omega_{\mathbf{q}\nu}$ is the phonon frequency. The total EPC strength is given by:

$$\lambda = 2 \int_0^\infty \frac{\alpha^2 F(\omega)}{\omega} d\omega. \tag{7}$$

The superconducting transition temperature $T_C$ can be calculated by:

$$T_C = \frac{\omega_{log}}{1.2} \exp\left[\frac{-1.04(1 + \lambda)}{\lambda - \mu^*(1 + 0.62\lambda)}\right]. \tag{8}$$

Here, $\omega_{log}$ is the logarithmically averaged phonon frequency, $\mu^*$ is the effective screened Coulomb parameter which is estimated from the DOS at the Fermi energy: $\mu^* \approx 0.26 N_F/(1 + N_F)$[7,21].

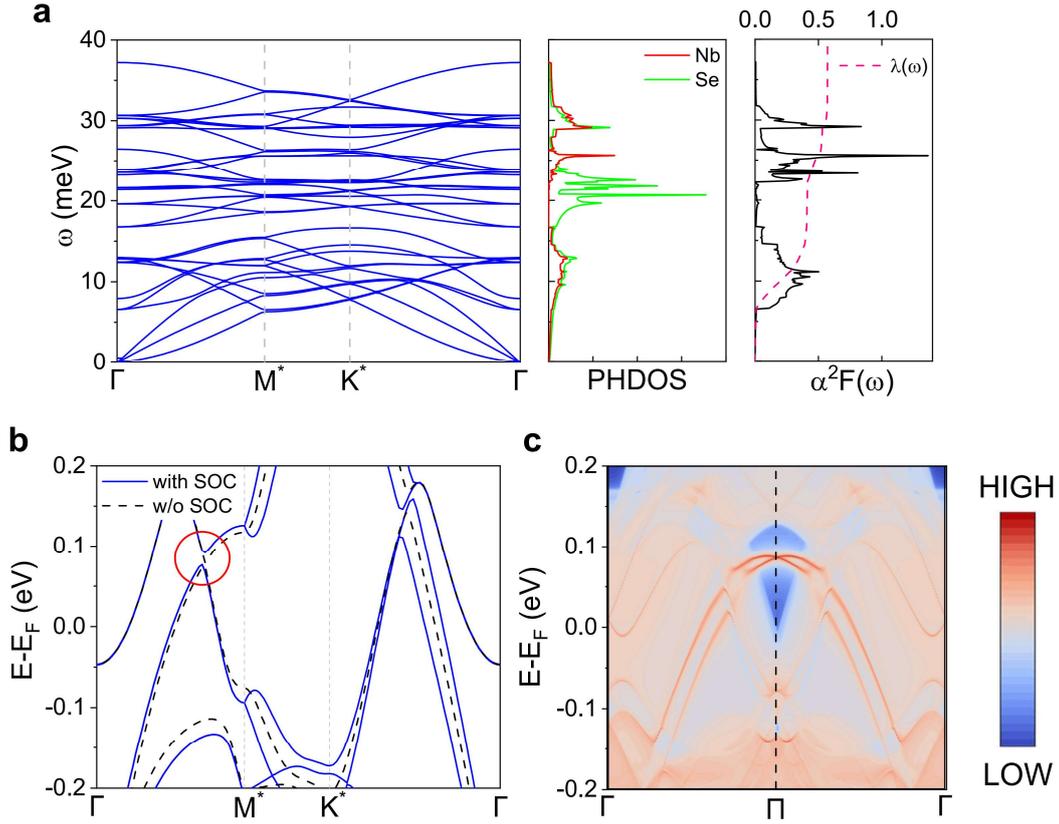

**Figure 4.** (a) The phonon spectrum, PHDOS, Eliashberg spectral function $\alpha^2F(\omega)$ and frequency-dependent integrated EPC $\lambda(\omega)$ of $2\times 2$ hollow CDW under 0.3 e⁻/f.u. doping concentration. (b) The band structure of $2\times 2$ hollow CDW under 0.3 e⁻/f.u. doping concentration with (blue solid lines) and without (black dashed lines) SOC. (c) The edge states of $2\times 2$ hollow CDW are projected on the (100) edge. Red colors indicate higher local DOS and blue colors represent the bulk band gap. The high-symmetry points are indicated in Figure S1.

Figures 4a and S3a show the calculated phonon spectra for $2\times 2$ hollow CDW and $2\times 2$ filled CDW structures, respectively, under 0.3 e⁻/f.u. doping concentration. The absence of imaginary phonon frequency indicates the stability of these two structures. And the small energy difference (~0.8 meV/f.u., see Figure 1f) between these two structures suggests that both may exist in experiments. As shown in Figure 4a and Supplementary Figure 3a, the vibration of Se atoms is concentrated in the middle-frequency region (i.e., 17meV to 22meV), while Nb and Se atoms have similar contributions to the phonon DOS (PHDOS) outside this region. However, the high PHDOS of Se atoms in the middle-frequency region has small contributions to $\alpha^2 F(\omega)$ as well as $\lambda$. The total EPC strength $\lambda$ is calculated to be 0.57 (0.54) for the $2 \times 2$ hollow ($2 \times 2$ filled) CDW phase. Considering the effective screened Coulomb parameter $\mu^*$ of 0.15, the $T_C$ is estimated to be 1.7 K and 1.3 K for $2 \times 2$ hollow CDW and filled CDW, respectively, both of which are smaller than the $T_C \sim 3$ K for the $3 \times 3$ CDW[14,16]. Thus, the superconductivity of NbSe$_2$ is suppressed under 0.3 e$^-$/f.u. doping concentration, which is consistent with the previous experimental results of the electron-doped 2D NbSe$_2$[17,21–23]. The main reason for this suppression is due to the reduced $\lambda$ of the $2 \times 2$ CDW systems under electron doping[8].

Figure 4b shows the electronic band structure of $2 \times 2$ hollow CDW of NbSe$_2$ with and without spin-orbital coupling (SOC). Without SOC, there is a Weyl node protected by the mirror symmetry on the $\Gamma M^*$ path. The SOC opens a band gap of about 15 meV (see the red circle in Figure 4b), resulting in the nontrivial topological phase characterized by a nonzero topological invariant Z$_2$ =1. According to the bulk-edge correspondence, the nonzero Z$_2$ invariant is closely related to the nontrivial helical edge states that emerge inside the gap between two energy bands in a semi-infinite system. As presented in Figure 4c, there are Dirac cone edge states connecting the valence and conduction bands. Therefore, the $2 \times 2$ hollow CDW phase is a 2D superconductor with clearly visible topological edge states near the Fermi level, which is rarely reported in 2D systems and may have promising applications in fault-tolerant quantum computation[33–35].

**Conclusion**

In summary, using first-principles calculations and mean-field theory, we construct a theoretical framework to study the mechanism of CDW formation and CDW transition in NbSe$_2$ from both phononic and electronic perspectives. We demonstrate that **q**-EPC makes the dominant contribution to the $3 \times 3$ CDW formation of NbSe$_2$ as well as the transition to $2 \times 2$ CDW under electron doping. In addition, we find that electron doping suppresses both the **q**-EPC $|g_\mathbf{q}|^2$ and the total EPC strength $\lambda$, leading to the simultaneous suppression of CDW and superconducting orders. All these results agree with previous works. More importantly, to the best of our knowledge, the $2 \times 2$ hollow CDW phase of monolayer NbSe$_2$ under 0.3 e$^-$/f.u. doping is the first reported topological superconductor driven by CDW transition. Such phenomenon is also found in monolayer TaSe$_2$. Our seminal work not only deepens the understanding of CDW theory, but also presents a new strategy for developing topological superconductor.

ACKNOWLEDGMENTS

C.C. thanks for the support from Shuimu Tsinghua Scholar Program. This research project was funded by the Ministry of Education, Singapore, under its MOE AcRF Tier 3 Award MOE2018-T3-1-002, and MOE AcRF Tier 2 Award MOE2019-T2-2-030. Computations were supported by the Center of Advanced 2D Materials (CA2DM) HPC infrastructure.

COMPETING INTERESTS

The authors declare no competing financial or non-financial interests.